# Sub-terahertz sound excitation and detection by long Josephson junction

Valery P. Koshelets[*]

*Kotel'nikov Institute of Radio-Engineering and Electronics (IREE), Russian Academy of Sciences*
*Moscow, 125009, Russia*

(Dated: December 07, 2013)

The paper reports on experimental observation of sub-terahertz sound wave generation and detection by a long Josephson junction. This effect was discovered at spectral measurements of sub-terahertz electromagnetic emission from a flux-flow oscillator (FFO) deposited on optically polished Si substrate. The "back action" of the acoustic waves generated by the FFO and reflected by the bottom surface of the Si substrate results in the appearance of resonant steps in the FFO IVCs with spacing as small as 29 nV for 0.3 mm substrate thickness; these steps manifest themselves in a pronounced resonant structure in the emission spectra with spacing of about 14 MHz precisely according to the Josephson relation. The mechanism of acoustic wave generation and detection by the FFO is discussed; a possibility for employing the discovered effect for FFO frequency stabilization has been demonstrated. A simple and reliable way to suppress the superfine resonant structure has been developed and proven; this invention allows for continuous frequency tuning and FFO phase locking at *any* desired frequency, all of which is vitally important for most applications.

## I. INTRODUCTION

Interaction between phonons and electrons has attracted the attention of researchers over the years. Attempts to observe the influence of coherent microwave phonons in the current-voltage characteristics of a superconducting tunnel junction (STJ) were first undertaken in 1965 [1]. The possibility of using STJ for non-equilibrium phonon generation and detection at frequencies above 100 GHz was experimentally demonstrated a few years later [2, 3]; the phonons originate on relaxation and recombination of excited quasiparticles. At low temperatures, reduced scattering allows acoustic phonons to propagate in a substrate over long distances, providing the possibility of using phonons at low temperature as a tool for solid state physics and phonon spectroscopy; see reviews [4, 5]. The frequency resolution of the relaxation phonon spectrometer is determined ultimately by the sharpness of the energy gap; values of about a few GHz have been achieved [4, 5].

Much greater resolution has been demonstrated by the use of AC Josephson current for phonon generation in superconducting tunnel junctions [6, 7]; in these papers a few possible mechanisms for phonon generation were considered. Direct generation of acoustic waves by the AC-Josephson oscillations occurs when the tunnel barrier is piezoelectric. On the other hand, a need for an alternative explanation was suggested in ref. [7] because the effect was observed in junctions with an amorphous barrier and amorphous materials are typically not piezoelectric (although some materials may become piezoelectric in the amorphous state because amorphization can remove the inversion symmetry [8]). Therefore, it was suggested that in the disordered material the AC electric field may instead act on uncompensated static charges with a finite dipole moment [7, 9, 10], resulting in coherent generation of acoustic waves.

The coupling strength of this process depends on the oxide properties; it can be comparable (or even be well above) to Werthamer processes [11], which basically are absorption of AC Josephson radiation energy by the quasiparticles and consequent photon assisted tunneling - the so called effect of Josephson self-coupling (JSC) [11-13].

The reverse effect - phonon induced increase in the critical current of Josephson junctions [14] and appearance of the constant-voltage steps in the IVCs of the SNS junctions [15] - has been observed experimentally under low frequency phonon excitation. Interaction of the AC Josephson current and phonons was found also for intrinsic $HT_c$ Josephson junctions ($Bi_2Sr_2CaCu_2O_8$ and $Tl_2Ba_2Ca_2Cu_3O_{10}$) [16 - 19]; it was observed as specific subgap structures in the form of current peaks (resonances) in the IVCs of these junctions. The obtained results have been explained [19 - 22] by the coupling between the intrinsic Josephson oscillations and phonons. Coupling to phonons was considered as a reason for decoherence of the superconductor quantum bits [23, 24].

Some time ago, quite unusual superfine resonance structure (SFRS) was observed [25] in the IVCs of the Nb-AlOx-Nb flux-flow oscillator (FFO); at that time no reasonable explanation was proposed. The FFO [26 - 28] is a long Josephson tunnel junction of overlap geometry in which an applied DC magnetic field and a DC bias current, $I_B$, drive a unidirectional flow of fluxons, each containing one magnetic flux quantum, $\Phi_0 = h/2e \approx 2*10^{-15}$ Wb. The symbol

$h$ is Planck's constant and $e$ is the elementary charge. An integrated control line with current $I_{CL}$ is used to generate the DC magnetic field applied to the FFO. According to the Josephson relation, the junction oscillates with a frequency $f = (1/\Phi_0)*V$ (about 483.6 GHz/mV) if it is biased at voltage $V$. The fluxons repel each other and form a chain that moves along the junction. The velocity and density of the fluxon chain and, thus, the power and frequency of the subTHz-wave signal emitted from the exit end of the junction due to the collision with the boundary may be adjusted independently by proper settings of $I_B$ and $I_{CL}$.

## II. EXPERIMENTAL SAMPLES AND TECHNIQUE

For comprehensive analysis of the superfine resonance structure, we studied Nb-AlO$_x$-Nb and Nb-AlN-NbN FFOs. The length, $L$, and the width, $W$, of the tunnel junctions used in our study are 400 μm and 16 μm, respectively. The value of the critical current density, $J_c$, is in the range 4 - 8 kA/cm$^2$ - giving a Josephson penetration depth, $\lambda_J$ = 6 - 4 μm. The active area of the FFO (i. e. the AlO$_x$ or the AlN tunnel barrier) is formed as a long window in the relatively thick (200 nm) SiO$_2$ insulation layer sandwiched between two superconducting films (base and wiring electrodes). The FFOs were fabricated from a high quality tri-layer structure [29] on the mono-crystalline silicon substrate of (001) orientation. We used commercially available double sided polished silicon wafers (room temperature resistivity > 10 kΩ*cm, thickness $d_S$ = 0.3 +/- 0.01 mm).

For wide-band measurements of the FFO spectra, a superconductor-insulator-superconductor (SIS) mixer has been integrated on the same chip with the FFO [30]; a simplified sketch of the device under the test is presented in the inset to Fig.1a. The FFO and the SIS junction are connected by a specially designed microstrip circuit that provides RF coupling in the range 300 - 800 GHz while the break at DC gives us a possibility to bias and to measure both devices independently; this circuit is presented schematically by a dashed line and the capacitor in the inset to Fig.1a. Due to the strong nonlinearity of the SIS mixer, it was utilized as a high-number harmonic mixer (HM) [30] in which the FFO signal under investigation beats with the n-th harmonic of an applied reference signal (of about 20 GHz) fed to the SIS mixer via coaxial cable from a synthesized signal generator. Signals at down-converted frequencies $f_{IF} = f_{FFO} - n \cdot f_{ref}$ can be analyzed using a conventional spectrum analyzer. The down-converted signals measured at the FFO frequencies ranging from 414 to 720 GHz are presented in Fig.1.

## III. RESULTS AND DISCUSSION

The spectrum of the frequency-locked FFO [30] operating at 720 GHz is shown in Fig.1a by the solid line. The spectrum recorded at fine FFO frequency tuning in the range 100 MHz is presented by the dash-dotted line. This spectrum was measured by using the so-called "Max Hold" regime when the maximum value in each spectral channel of the analyser (601 points per range) is memorized over long enough time, providing that the FFO frequency is tuned by fine adjustment of the bias or CL current. The amplitude of the down-converted signal is almost constant for the FFO frequency 720 GHz (dash-dotted line in Fig.1a); while at decreasing of the FFO frequency a well-defined resonant structure appeared in the down-converted spectra (Fig.1b).

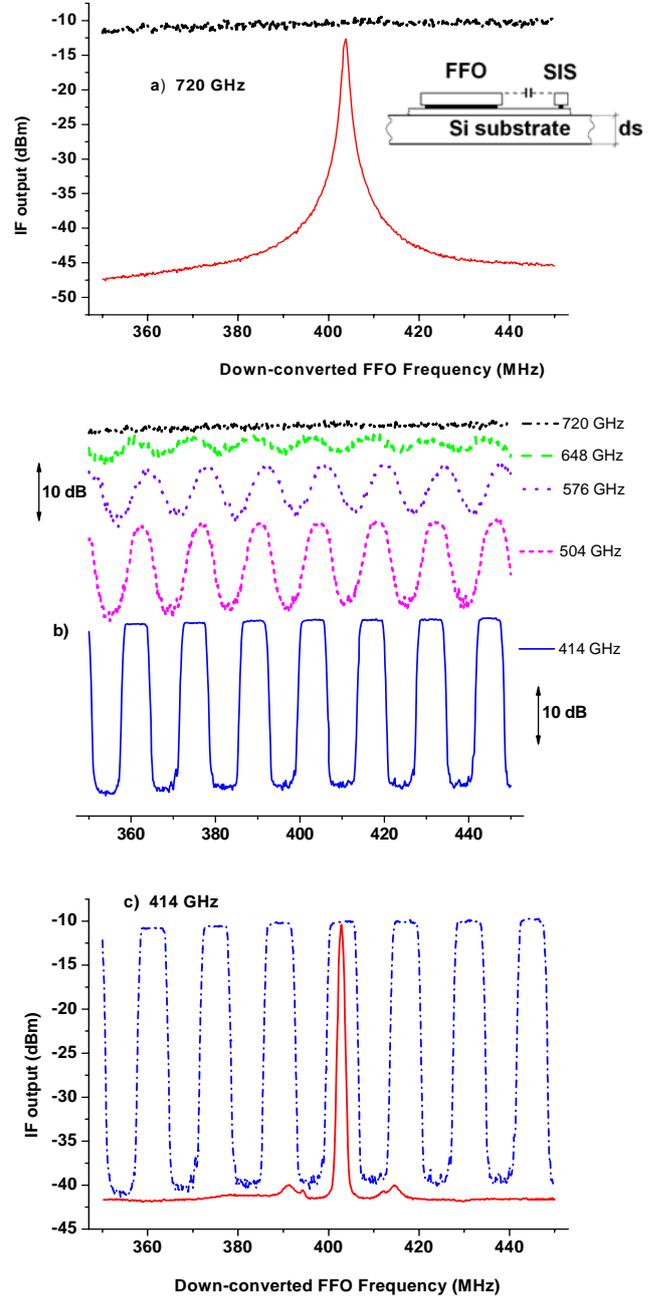

FIG. 1. (Color online) Down-converted spectra of the Nb-AlO$_x$-Nb FFO measured at different FFO frequencies at T= 4.2 K by spectrum analyzer in the regime "Max Hold" (see text) with Resolution Band Width (RBW) = 1 MHz at fine tuning of the bias current; the frequency-locked FFO spectra measured at FFO frequencies 720 and 414 GHz are presented by solid lines in graphs "a" and "c". Sketch of the experimental sample is shown as an inset to Fig.1a (see text).

At low FFO frequencies, the SFRS is very pronounced and down-converted power in the dips is at least 1000 times (30 dB) smaller than in the maximum (see Fig.1c); actually minimum level in this case is determined by the noise level of the HM. One can see that the FFO frequency can be continuously tuned only in a small range, while frequencies between these stable regions cannot be obtained. Even a small change of the bias current near the edge of the stable region results in a "jump" of the FFO voltage (frequency) to the next stable region. The distance between resonances is equal to 14.1 MHz; exactly the same resonance spacing was measured for all FFO frequencies and for all tested FFOs fabricated on the 0.3 mm thick silicon substrates described above.

Very similar behavior was measured also for the Nb-AlN-NbN junctions; the down-converted spectra recorded at the FFO frequency 396 GHz are presented in Fig.2a. Again, spacing between resonances is 14.1 MHz; note that in this case the spectra were measured for quite different type of tunnel barrier (AlN instead of $Al_2O_3$). By using a specially developed measuring procedure, the frequency readings from the spectrum analyzer were recorded simultaneously with bias current values at fine adjusting of the $I_b$ at constant $I_{CL}$. By using the Josephson relation, the IVC can be reconstructed with high precision (better than 1 nV), see Fig. 2a, b; at this low FFO frequency, the SFRS in the IVC is highly hysteretic. The measured SFRS is superimposed on the "regular" Fiske steps (FS); see Fig. 2b, c. Differential resistance on the SFRS is much lower than that measured on the Fiske step ($R_d^{SFRS}$ - as small as 0.0002 Ω, that is 36 times lower than the $R_d^{FS}$ = 0,0072 Ω, see Fig.2b), which results in further decrease of the FFO linewidth (down to values well below 100 kHz). Although a detailed study of the linewidth dependence on the FFO parameters [27] looks complicated due to considerable modification of the FFO line shape by the presence of the SFRS, extremely low values of the $R_d^{SFRS}$ results in additional stabilization of the FFO frequency that can be employed for many applications.

We attribute this superfine structure to the manifestation of resonant interaction of the acoustic waves with the Josephson oscillator. A few different mechanisms were proposed that may couple electron oscillations and phonons: (i) excitation of phonons in a tunnel barrier due to the electromagnetic interaction between the ionic charges and the charges of conduction electrons [6, 7, 9, 10, 18, 21, 22], or via the ac-Josephson effect in a tunnel barrier made of a piezo and ferroelectric materials; (ii) emission of phonons via non-equilibrium quasiparticle relaxation in the electrodes (not the barrier) caused by electron-phonon interaction [2, 4, 5, 19]; (iii) the dependence of the tunneling matrix element on lattice displacements [20, 31]. The mechanism for generation of phonons in Josephson junctions [31] is based on the excitation of the long wavelength acoustic resonance modes in the dielectric layer of the contact, which can influence the shape of the IVC of the junction in the same way as excitation of electromagnetic cavity modes. This approach was extended [20] by including in consideration *all* optical phonons in superconductors (not only in the intermediate dielectric layer). To distinguish between these mechanisms, additional research is required, but that is outside the scope of this paper.

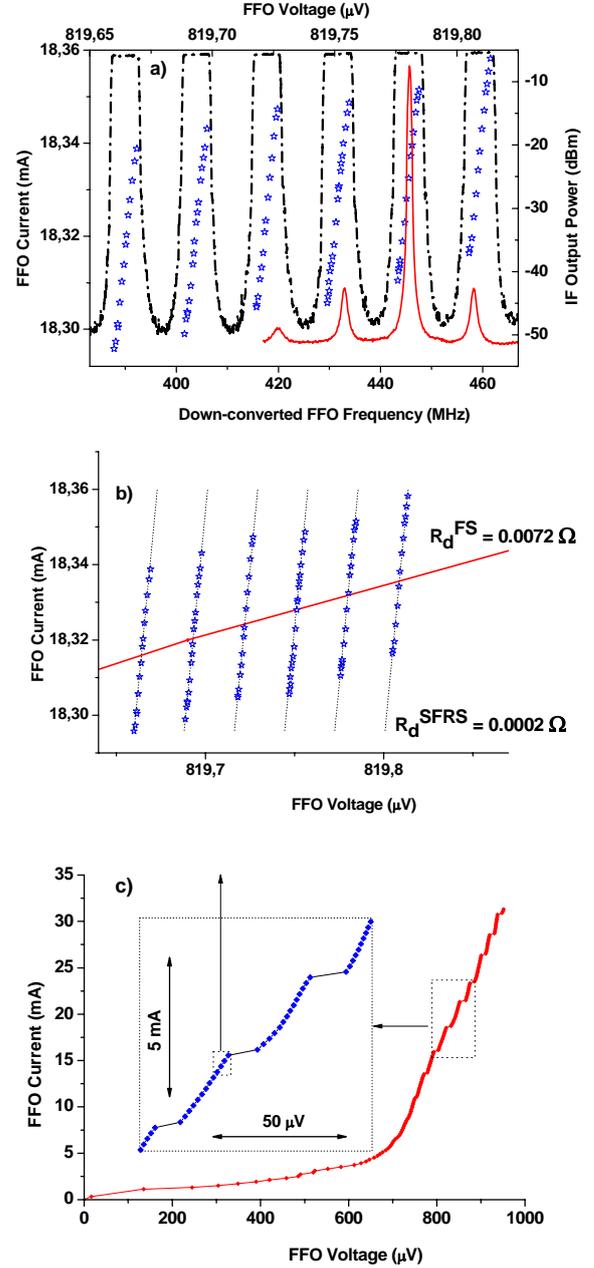

FIG. 2. (Color online) Down-converted spectra and IVCs of the Nb-AlN-NbN tunnel junction at T= 4.2 K measured at constant magnetic field produced by control line at $I_{cl}$ = 31.5 mA (FFO frequency = 396 GHz): a) spectrum recorded in the regime "Max Hold" at fine tuning of the bias current (dotted line), the FFO spectrum at constant FFO bias (solid line) and a small part of the FFO IVC reconstructed from the frequency reading by Josephson relation (open symbols – left and top axes); b) reconstructed FFO IVC on the top of the "averaged" Fiske step shown by solid line; c) the FFO IVCs measured by traditional DC electronics.

According to our explanation, the experimentally measured IVC are caused by excitation of the standing acoustic waves in the Si substrate. It is known that a considerable part of the power emitted by the FFO is reflected back; at low damping, these oscillations may reach the entry end of the FFO, raising the standing waves that manifest themselves as the Fiske steps. The standing electromagnetic waves of large amplitude (existing at least on the emitting end of the junction) excite acoustic waves. Note that even at higher voltages ($V > V_g/3$, $V_g$ is the gap voltage of the FFO), where the Fiske steps could be suppressed [13] due to higher dumping caused by the JSC effect, the standing electromagnetic waves still exist at the emitting end of the junction and can excite acoustic waves of considerable amplitude.

Since flatness (parallelism) of the Si substrates is quite good (thickness variation is well below 5 μm over 100 mm wafer size), the acoustic wave is reflected back to the point of emission with accuracy much better 0.1 μm. At frequencies where the Si substrate thickness is equal to the integer number of acoustic wave-lengths, the reflected wave will arrive in phase with electromagnetic oscillations, resulting in an increase of the current amplitude, while in between these resonances the oscillations will be suppressed.

At least two experimental facts lead to such a conclusion. Firstly, the frequency distance between two adjacent resonances coincides with the distance between sound resonances in the silicon substrate:

$$\Delta f = V_L/d_S*2, \qquad (1)$$

where $V_L$ is longitudinal speed of sound in Si along [001] direction. For $V_L$ = 8480 m/s [32] and $d_S$ = 0.3 mm, the calculated resonance spacing is 14.1 MHz, which precisely corresponds to experimentally measured data (see Fig. 1, 2). The frequency distance between two adjacent resonances is inversely proportional to the substrate thickness $d_S$; for the FFOs fabricated on a thicker Si substrate ($d_S$ = 0,525 mm), the resonance spacing was about 8 MHz (see Fig. 3) that again corresponds well to the calculated value (1).

Secondly, after treatment of the opposite (bottom) substrate surface with an abrasive powder, the superfine resonant structure completely disappeared (see Fig. 4). We used a set of powders with particle size from 1 to 10 μm, resulting in a Root Mean Square (RMS) surface roughness measured by Atomic Force Microscope (AFM) from 30 to 280 nm, which is well above wave-length of the acoustic wave at 500 GHz of about 13 nm (note that for polished Si surface the RMS is of about 0.1 nm, see e.g. [33]. It seems that the acoustic waves reflected from rough Si surface arrive at the FFO plane in arbitrary phase; that makes establishment of standing acoustic waves impossible giving us a possibility to phase lock the FFO [34] at *any* desirable frequency that is vitally important for most practical applications. It was found that chemical etching of the bottom surface of the Si substrate (RMS roughness of about 250 nm) also completely eliminates the appearance of the SFRS (see Fig. 5). Such Si substrates with a chemically etched bottom surface are commercially available and were used for fabrication of an integrated receiver with phased-locked FFO [28] that was successfully implemented for atmosphere monitoring from a high-altitude balloon [35]. On the other hand, the roughness of the etched Si substrate is negligibly small at sub-THz frequencies and allows good RF coupling of the integrated receiver that is installed on the flat surface of the synthesized elliptical Si lens [35].

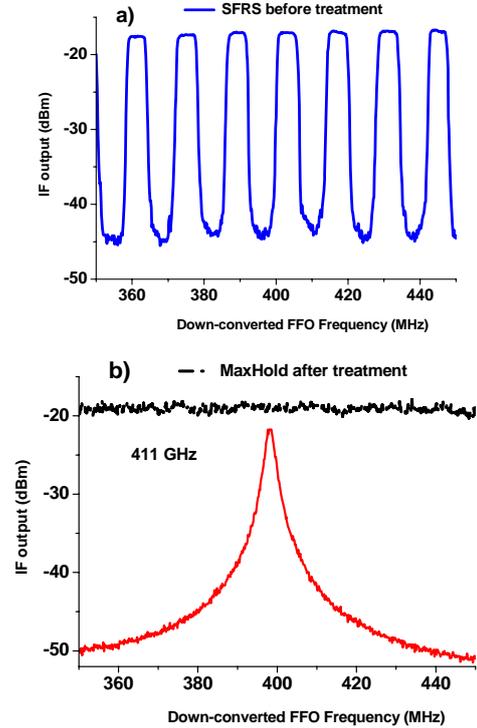

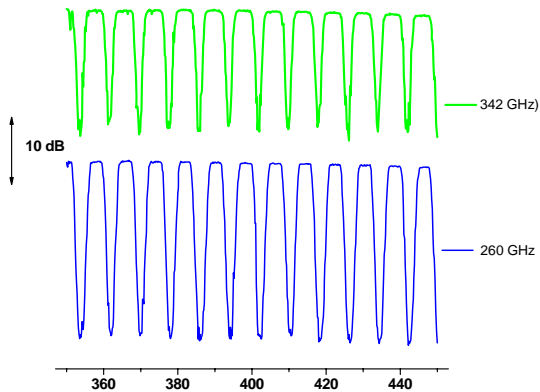

FIG. 3. (Color online) Down-converted spectra of the Nb-AlO$_x$-Nb FFO fabricated on the 0.525 mm thick Si substrate, measured at different FFO frequencies at T= 4.2 K by spectrum analyzer in the regime "Max Hold" (see text) at fine tuning of the bias current.

FIG. 4. (Color online) Down-converted spectra of the Nb-AlO$_x$-Nb FFO fabricated on the 0.3 mm thick Si substrate before (a) and after abrasive treatment (b), measured at different FFO frequencies at T= 4.2 K by spectrum analyzer in the regime "Max Hold" (see text); the frequency-locked FFO spectrum measured at FFO frequency 411 GHz is presented by solid lines in graph "b".

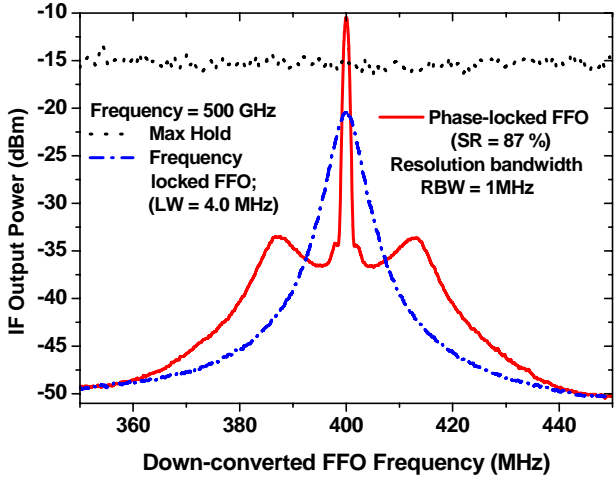

FIG. 5. (Color online) Down-converted spectra of the Nb-AlN-NbN FFO fabricated on the 0.525 mm thick Si substrate with chemically etched bottom surface, measured by spectrum analyzer at the FFO frequency 500 GHz in the regime "Max Hold" (see text) at fine tuning of the bias current; the frequency-locked and phase-locked FFO spectra measured with resolution band width 1 MHz are presented by dash and solid lines respectively.

This explanation of the SFRS was confirmed by preliminary theoretical consideration [36] where coherent phonon radiation and detection due to the interaction of Josephson's electromagnetic oscillations with mechanical displacement field have been analyzed. The Josephson tunneling structure together with silicon substrate constitutes a high overtone composite resonator for bulk acoustic waves propagating normally to the layers. Resonant generation of coherent acoustic waves revealed itself as a superfine structure in IVC, similar to the usual Fiske steps caused by reflection of electromagnetic waves in the junction resonance cavity.

A difference between values of the down-converted power measured in the dips and in the maximum characterizes the "magnitude" of the SFRS effect. To compare quantitatively the strength of the effect at different conditions, the values measured from the down-converted spectrum (see Fig. 1 - 4) should be normalized to the maximum possible depth value. This quantity was determined at each point from experimentally measured FFO spectra taking into account maximal amplitude of the down-converted signal, the FFO linewidth, the SFRS spacing and the HM noise level. The presence of the SFRS considerably modifies the shape of the FFO line that is Lorentzian [27] for the FFO in the absence of SFRS (see Fig. 1a, 4b, 5). The resulting line shape depends on the relation between initial FFO linewidth and the SFRS spacing.

Dependence of the normalized SFRS value on the FFO frequency is presented in Fig. 6 both for the Nb-AlOx-Nb and the Nb-AlN-NbN FFOs. The difference between these two FFO types presumably can be explained by greater piezoelectricity in the AlN barrier and/or smaller losses for the Josephson oscillations at high frequencies in the Nb-AlN-NbN FFO, with the result that higher RF power is available for excitation of acoustic resonances. This explanation is supported by data measured for both types of the FFO at temperatures of about 5 K (solid symbols in Fig. 6), as well as by dependencies of the SFRS on the bias current (data for the Nb-AlOx-Nb FFO at T = 4.2 K are presented in Fig. 7). It should be noted that no change in the normalized SFRS value was found at the crossing of the "boundary" voltage for the JSC effect $V_b = V_g/3$ ($V_g$ is the gap voltage of the FFO) where the abrupt merge of Fiske steps caused by an increase of the internal damping in the long junction due to quasi-particle tunneling takes place [13]; this voltage corresponds to the FFO frequency 450 and 610 GHz for the Nb-AlOx-Nb and the Nb-AlN-NbN circuits respectively.

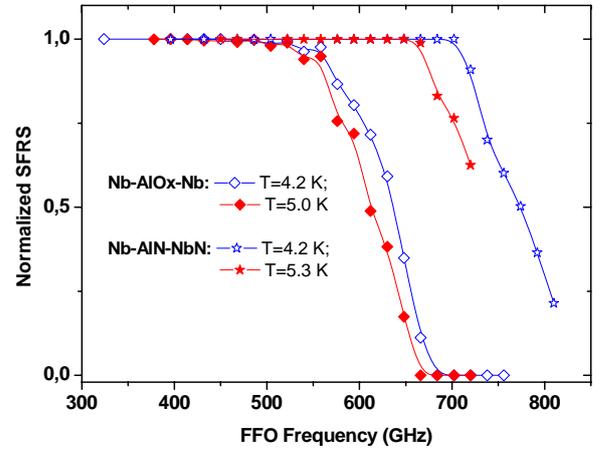

FIG. 6. (Color online) Normalized SFRS depth on the FFO frequency for the Nb-AlOx-Nb and Nb-AlN-NbN FFOs at different temperatures. Points are connected by lines as a guide for the eye.

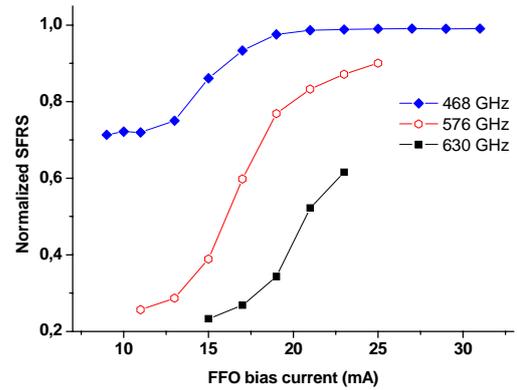

FIG. 7. (Color online) Dependencies of the SFRS on the bias current for the Nb-AlOx-Nb FFO at T= 4.2 K, normalized on maximum possible at the each operation point. Points are connected by lines as a guide for the eye.

## IV. CONCLUSIONS

An ability of the Josephson junction to generate and to detect the coherent acoustic waves has been demonstrated experimentally at frequencies up to 800 GHz. Frequency resolution well below 1 MHz can be realized for the frequency (or phase-locked) FFO and that opens new possibilities for solid state physics research and phonon spectroscopy. The superfine resonant structure in the FFO IVCs is attributed to acoustic wave generation by the FFO and excitation of acoustic wave resonances in thick Si substrate. The SFRS effect can be avoided by proper treatment of the bottom surface; on the other hand, this effect can be employed for high-resolution phonon spectroscopy at well-defined frequencies without additional FFO locking.

### ACNOWLEGEMENTS

The author thanks Vladimir Krasnov, Vladislav Kurin, Georgy Mansfeld, Jesper Mygind, Natalia Polzikova, Valery Ryazanov and Alexey Ustinov for fruitful discussions; Pavel Dmitriev for fabrication of the experimental samples; Yuri Tokpanov for measurements of the surface roughness and valuable comments. The work was supported by the RFBR and the Ministry of Education and Science of the Russian Federation.